\documentclass[10pt,twocolumn]{article}
\usepackage[a4paper,top=2.1cm,bottom=2.1cm,left=1.85cm,right=1.85cm,columnsep=0.62cm]{geometry}
\usepackage{amsmath,amssymb,bm}
\usepackage{graphicx}
\usepackage{booktabs}
\usepackage{multirow}
\usepackage[raggedright]{titlesec}
\usepackage[font=small,labelfont=bf]{caption}
\usepackage{xcolor}
\usepackage[colorlinks=true,allcolors=blue]{hyperref}
\usepackage{dblfloatfix}
\newcommand{\Bcr}{B_{\rm cr}}
\newcommand{\dn}{\Delta n}
\newcommand{\kp}{\kappa_{p}}
\newcommand{\ks}{\kappa_{s}}
\newcommand{\gs}{\gamma_{s}}
\newcommand{\Npar}{N_{\parallel}}
\newcommand{\Nperp}{N_{\perp}}
\newcommand{\RNS}{R_{\rm NS}}
\newcommand{\Aang}{A}
\newcommand{\dmax}{\epsilon_{\rm max}}
\newcommand{\etaP}{\eta_{p}}
\newcommand{\etaS}{\eta_{s}}

\begin{document}
\twocolumn[
\begin{@twocolumnfalse}
{\centering
\vspace*{2mm}
{\bfseries\LARGE Angular structure of finite-field vacuum birefringence in magnetars\par}
\vspace{5mm}
{\large S.~Abbassi\textsuperscript{1,a}\quad S.~R.~Valluri\textsuperscript{1,b}\par}
\vspace{3mm}
{\small\textsuperscript{1}\,Department of Physics and Astronomy, The University of Western Ontario, London, ON N6A 3K7, Canada\par}
\vspace{2mm}
{\small Received: date / Accepted: date\par}
\vspace{1mm}
\vspace{5mm}}
\begin{abstract}\noindent
Vacuum birefringence in a strong magnetic field is often modelled by combining the finite-field splitting for perpendicular propagation with the familiar $\sin^2\theta$ angular factor. That separation is exact only at leading response order. Continuing our previous study of the field-strength dependence of the birefringence, we examine the angular dependence of the full one-loop refractive indices in the local Heisenberg--Euler theory. We show analytically that, on the positive-birefringence branch, the usual factorized form systematically overestimates the exact splitting away from perpendicular propagation. The relative correction is largest close to the magnetic-field direction, where the absolute birefringence itself vanishes. Compact analytic expansions expose the sign, scaling, and angular location of the correction throughout the magnetar-relevant range $B/B_{\rm cr}\le100$, while an independent proper-time calculation verifies the underlying one-loop coefficients. In centred-dipole applications the correction is sub-percent for moderate magnetar fields but reaches several percent for the strongest sources; surface averaging and radial phase accumulation reduce its physical weight as the field decreases away from the stellar surface. The result separates finite-field amplitude and angular corrections and provides controlled analytic benchmarks for precision polarization transport in strongly magnetized environments.
\end{abstract}
\vspace{2mm}
{\footnotesize\noindent\textsuperscript{a}\,e-mail: sabbassi@uwo.ca (corresponding author)\\
\textsuperscript{b}\,e-mail: valluri@uwo.ca\par}
\vspace{5mm}
\end{@twocolumnfalse}]

\section{Introduction}\label{sec:intro}
Nonlinear photon propagation is one of the characteristic low-energy consequences of quantum electrodynamics. In the Heisenberg--Euler description, virtual electron--positron fluctuations make the vacuum respond as a nonlinear electromagnetic medium \cite{HeisenbergEuler1936,Schwinger1951}. A magnetic background then selects a preferred direction, and the two physical photon eigenmodes acquire different dispersion relations and refractive indices \cite{Adler1971,TsaiErber1975}. The resulting vacuum birefringence is extremely small in ordinary laboratory magnetic fields, but it becomes progressively more important as the field approaches the QED critical scale $\Bcr=m_e^2/e\simeq4.4\times10^{13}$~G.

Magnetars provide the most direct astrophysical route to this regime. Their inferred surface fields are typically $10^{14}$--$10^{15}$~G, so photons emerging from the stellar atmosphere and propagating through the inner magnetosphere can sample fields comparable to, or well above, $\Bcr$. In that environment vacuum birefringence influences the evolution of the polarization eigenmodes and therefore enters calculations of X-ray polarization transport. The growing body of X-ray polarimetry, including recent phase- and energy-resolved observations of 1E~1547.0$-$5408, has made accurate modelling of this propagation increasingly relevant \cite{StewartNature2026,DinhThi2026}. Such measurements do not determine a single QED coefficient in isolation: their interpretation also depends on the atmosphere or surface-emission model, plasma response, magnetic geometry, and viewing configuration. They nevertheless provide a strong physical motivation for controlling the vacuum contribution as accurately as the adopted approximation permits.

In the weak-field, soft-photon limit the refractive-index splitting takes the familiar Cotton--Mouton form
\begin{equation}
\dn_{\rm wf}=\frac{\alpha}{30\pi}\,\xi^2\sin^2\theta,
\qquad \xi\equiv\frac{B}{\Bcr},
\label{eq:cm}
\end{equation}
where $\theta$ is the angle between the photon wave vector and the external magnetic field. Equation~(\ref{eq:cm}) combines two physically distinct simplifications. The first is the weak-field expansion of the QED response; the second is the separation of the angular dependence into a pure $\sin^2\theta$ factor. Once the field becomes supercritical, there is no reason for these two approximations to remain equally accurate.

The finite-field response of the magnetized QED vacuum has been developed from both effective-action and polarization-tensor approaches \cite{TsaiErber1975,HeylHernquist1997,VillalbaShabad2012,HattoriItakuraI,HattoriItakuraII}. In particular, Valluri et al. expressed the relevant one-loop Heisenberg--Euler derivatives in terms of Hurwitz-zeta and digamma functions and applied the resulting strong-field refractive indices to neutron-star vacuum birefringence \cite{Valluri2017}. Finite-field vacuum polarizabilities are also standard ingredients of magnetized neutron-star atmosphere and transport calculations \cite{Potekhin2004,vAL2006}. These studies establish the underlying anisotropic dispersion; the question addressed here is narrower. We ask how accurately the exact arbitrary-angle splitting can be represented by the perpendicular-propagation result multiplied by $\sin^2\theta$ once the full field dependence of the one-loop coefficients is retained.

This paper is a direct continuation of Paper~I \cite{PaperI}, our previous finite-field study of vacuum birefringence and magnetar polarization transport. In that work we kept the finite-field one-loop response and the refractive-index normalization $\gamma_s$ unexpanded, and quantified how the weak-field amplitude of the birefringence fails at magnetar-strength fields. The arbitrary-angle indices were already part of that framework, but the angular factorization itself was not isolated. The present paper addresses that remaining step: after the finite-field amplitude has been treated consistently, what correction is introduced by retaining the full angular dependence of the refractive indices rather than imposing the weak-response $\sin^2\theta$ form?

The point is not simply to re-evaluate the same indices at more angles. Their common local constitutive form permits an exact positive-series representation for the angular dependence. From that representation the sign, monotonicity, and location of the maximum factorization error follow analytically, independently of a numerical scan or of the special-function representation used to evaluate the one-loop coefficients. The finite-field amplitude studied in Paper~I and the angular non-separability studied here are therefore distinct pieces of the strong-field response.

The exact splitting still contains an overall $\sin^2\theta$ factor, but its coefficient becomes angle dependent at finite response. We derive compact expansions that expose the scaling of this effect and provide analytic benchmarks for numerical implementations, while the exact rationalized expression remains the preferred form when the response coefficients are already known. Because the largest relative correction occurs close to the magnetic-field direction, where the absolute splitting simultaneously tends to zero, we also carry the calculation into a centred-dipole magnetar geometry and quantify the hierarchy between local, surface-averaged, and radially accumulated phase corrections.

The paper is organized as follows. Section~\ref{sec:framework} defines the finite-field one-loop response and fixes the polarization and notation conventions. Section~\ref{sec:analytic} develops the angular dependence and its analytic approximations. Section~\ref{sec:accuracy} discusses the domain of validity, including higher-loop, momentum, and thermal effects. Section~\ref{sec:numerics} describes the numerical validation, and Sect.~\ref{sec:dipole} applies the result to magnetar-strength dipole fields. Section~\ref{sec:conclusion} summarizes the physical implications. We use natural units $c=\hbar=1$ except where dimensional scales are quoted.

\section{Finite-field vacuum response}\label{sec:framework}
We follow the notation and normalization introduced in Paper~I \cite{PaperI}. For a constant electromagnetic background the Lorentz invariants are
\begin{align}
\mathcal F&=\frac14F_{\mu\nu}F^{\mu\nu}=\frac12(B^2-E^2),\\
\mathcal G&=\frac14F^{*}_{\mu\nu}F^{\mu\nu}=-\mathbf E\!\cdot\!\mathbf B.
\label{eq:invariants}
\end{align}
The effective Lagrangian is written as $\mathcal L_{\rm eff}=-\mathcal F+\mathcal L^{(1)}$, where $\mathcal L^{(1)}$ is the one-loop Heisenberg--Euler contribution. We denote its derivatives by $\gamma_{\mathcal F}$, $\gamma_{\mathcal FF}$, and $\gamma_{\mathcal GG}$, evaluated for the purely magnetic background $\mathcal G=0$ and $\mathcal F=B^2/2$. The refractive-index normalization is therefore
\begin{equation}
\gs\equiv-\mathcal L_{\mathcal F}=1-\gamma_{\mathcal F}.
\label{eq:gs}
\end{equation}
This separation of the classical Maxwell term from the one-loop correction is the same convention used in Paper~I and reproduces the standard weak-field coefficients.

For compactness we write
\begin{equation}
B^2\gamma_{\mathcal GG}=\frac{\alpha}{2\pi}\Npar,
\qquad
B^2\gamma_{\mathcal FF}=\frac{\alpha}{2\pi}\Nperp,
\end{equation}
and define the dimensionless response parameters
\begin{equation}
\kp=\frac{B^2\gamma_{\mathcal GG}}{\gs},\qquad
\ks=\frac{B^2\gamma_{\mathcal FF}}{\gs}.
\label{eq:kappas}
\end{equation}
The special-function expressions for these derivatives are those given in Paper~I and are equivalent to the Hurwitz-zeta/digamma representation used by Valluri et al. \cite{Valluri2017}. Their field variable $h=B_{\rm cr}/(2B)$ is related to the present one by $h=(2\xi)^{-1}$.

The weak- and strong-field limits needed below are
\begin{align}
\Npar &\to \frac{14}{45}\xi^2, & \Nperp &\to \frac{8}{45}\xi^2, && \xi\ll1,\\
\Npar &=\frac23\xi-C_0+\mathcal O(\xi^{-1}\ln\xi), & \Nperp&\to\frac23, && \xi\gg1,
\label{eq:asym}
\end{align}
so that $\kp\sim\alpha\xi/(3\pi)$ while $\ks\to\alpha/(3\pi)$ at strong field.

The two physical photon modes are labelled as in Paper~I and in the Adler convention: $\parallel$ denotes the mode whose photon electric field lies in the $(\mathbf B,\mathbf k)$ plane, whereas $\perp$ denotes the mode with electric field perpendicular to that plane. Some strong-field papers instead label the modes using the photon magnetic field; the labels are then interchanged, although the physical eigenvalues are the same \cite{Valluri2017}. With the present convention, the low-momentum refractive indices at propagation angle $\theta$ are \cite{VillalbaShabad2012,PaperI}
\begin{align}
 n_{\perp}&=\bigl(1-\ks\sin^2\theta\bigr)^{-1/2},\label{eq:nperp}\\
 n_{\parallel}&=\left(\frac{1+\kp}{1+\kp\cos^2\theta}\right)^{1/2}.
\label{eq:npar}
\end{align}
Both indices tend to unity for propagation along the field, where the birefringence vanishes. Keeping only the leading response gives
\begin{equation}
\dn\equiv n_{\parallel}-n_{\perp}
=\frac12(\kp-\ks)\sin^2\theta+\mathcal O(\kappa^2),
\label{eq:leading}
\end{equation}
Thus the familiar $\sin^2\theta$ law follows at leading response order even when the field dependence of $\kp$ and $\ks$ is retained. The correction studied below is generated by the nonlinear dependence of the refractive indices on these same finite-field response coefficients.

\section{Angular structure of the birefringence}\label{sec:analytic}
\subsection{Exact angular form}
Set
\begin{equation}
u\equiv\sin^2\theta,
\qquad
\etaP\equiv\frac{\kp}{1+\kp},
\qquad
\etaS\equiv\ks.
\label{eq:etas}
\end{equation}
Equations~(\ref{eq:nperp})--(\ref{eq:npar}) then become
\begin{equation}
 n_{\parallel}=(1-\etaP u)^{-1/2},
 \qquad n_{\perp}=(1-\etaS u)^{-1/2}.
\label{eq:common}
\end{equation}
This form makes the finite-field angular coupling transparent. Rationalizing the difference also gives a numerically stable expression for the exact splitting,
\begin{equation}
\dn(u)=\frac{(\etaP-\etaS)u}
{(1-\etaP u)(1-\etaS u)\,[n_{\parallel}(u)+n_{\perp}(u)]}.
\label{eq:rational}
\end{equation}
The factor $u=\sin^2\theta$ is therefore exact. What ceases to be exact at finite response is the additional assumption that the coefficient multiplying $u$ is angle independent.

Using
\begin{equation}
(1-x)^{-1/2}=\sum_{m=0}^{\infty}c_mx^m,
\qquad
c_m=\frac{1}{4^m}{2m\choose m}>0,
\end{equation}
we obtain
\begin{equation}
\frac{\dn(u)}{u}=\sum_{m=1}^{\infty}c_m
\bigl(\etaP^m-\etaS^m\bigr)u^{m-1}.
\label{eq:series}
\end{equation}
On the positive-birefringence branch considered here, $0\le\etaS<\etaP<1$ over the entire numerical interval. Every coefficient in Eq.~(\ref{eq:series}) is therefore positive, and
\begin{equation}
\frac{{\rm d}}{{\rm d}u}\left(\frac{\dn}{u}\right)>0,
\qquad 0<u<1.
\label{eq:monotonic}
\end{equation}
The direction of the correction is therefore fixed analytically, without relying on a numerical scan. Defining the normalized angular correction factor
\begin{equation}
\Aang(\xi,\theta)\equiv
\frac{\dn(\xi,\theta)}{\dn(\xi,\pi/2)\sin^2\theta},
\label{eq:Fdef}
\end{equation}
we find that $\Aang$ increases monotonically with $u$ from its near-axis limit to unity at perpendicular propagation. Hence
\begin{equation}
\Aang(\xi,0)\le\Aang(\xi,\theta)\le1,
\label{eq:Fbound}
\end{equation}
and, on the positive-birefringence branch considered here, the factorized prescription necessarily overestimates the exact finite-field splitting for $0<\theta<\pi/2$. The relative error is largest as $\theta\to0$, even though the absolute splitting vanishes in the same limit.

\subsection{Analytic approximations}
The exact near-axis limit follows from Eq.~(\ref{eq:rational}),
\begin{equation}
\Aang(\xi,0)=
\frac{\tfrac12[\kp/(1+\kp)-\ks]}
{\sqrt{1+\kp}-(1-\ks)^{-1/2}}.
\label{eq:Fzero}
\end{equation}
Expanding the full ratio through first relative order gives a result valid at arbitrary angle,
\begin{equation}
\Aang_1(\xi,\theta)=1-\frac{3}{4}(\kp+\ks)\cos^2\theta.
\label{eq:F1}
\end{equation}
The first finite-field correction is strongest for propagation close to the magnetic-field direction and vanishes continuously for perpendicular propagation. The near-axis limit follows directly by setting $\theta=0$ in Eq.~(\ref{eq:F1}).

Carrying the ratio to second relative order yields
\begin{equation}
\begin{aligned}
\Aang_2(\xi,\theta)=1&-\frac{3}{4}(\kp+\ks)c^2 \\
&+\frac{c^2}{16}\Big[(1+10c^2)\kp^2
+(-2+10c^2)\kp\ks \\
&\hspace{23mm}+(-11+10c^2)\ks^2\Big],
\end{aligned}
\label{eq:F2}
\end{equation}
where $c\equiv\cos\theta$ and the omitted terms are $\mathcal O(\kappa^3)$. Equation~(\ref{eq:F2}) satisfies $\Aang_2(\pi/2)=1$ identically and retains the exact location of the maximum error.

For consistency with the surface and phase diagnostics below, we measure the factorization error relative to the exact splitting,
\begin{equation}
\epsilon(\xi,\theta)\equiv
\frac{\dn_{\rm fact}-\dn_{\rm exact}}{\dn_{\rm exact}}
=\Aang^{-1}(\xi,\theta)-1,
\label{eq:epsilon}
\end{equation}
which is non-negative on the branch considered here. Its maximum is the near-axis limit,
\begin{equation}
\dmax(\xi)\equiv\Aang^{-1}(\xi,0)-1.
\label{eq:dmax}
\end{equation}
At leading order,
\begin{equation}
\dmax\simeq\frac{3}{4}(\kp+\ks)
\xrightarrow[\xi\gg1]{\kp\ll1}\frac{\alpha}{4\pi}\xi.
\label{eq:dmaxasym}
\end{equation}
The last relation is an overlap asymptote, requiring both the strong-field form of the response and $\kp\ll1$; it should not be read as an unrestricted $\xi\to\infty$ limit. Once $\kp$ is no longer small, the unexpanded one-loop algebra is no longer a strict perturbative ordering.

\begin{figure*}[!t]
\centering
\includegraphics[width=0.985\textwidth]{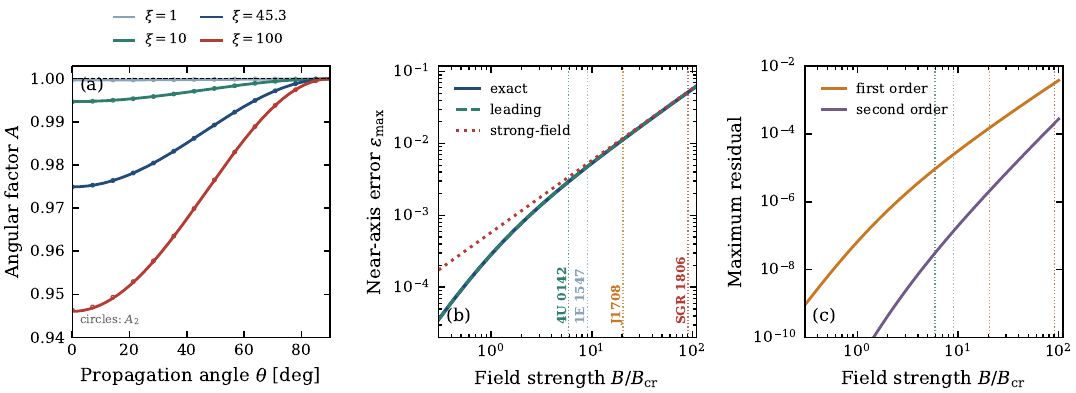}
\caption{Analytic angular structure. \textbf{(a)} Exact factorization ratio $\Aang$ for representative fields; open circles show the second-order expansion $\Aang_2$ from Eq.~(\ref{eq:F2}). \textbf{(b)} Exact-normalized near-axis error $\dmax$ compared with the leading expression $3(\kp+\ks)/4$ and its strong-field overlap asymptote. Vertical dotted lines mark the polar fields $B_p=2B_d$ adopted for the illustrative magnetars in Table~\ref{tab:sources}. \textbf{(c)} Maximum absolute residual over angle for $\Aang_1$ and $\Aang_2$. The second-order expansion remains within $2.72\times10^{-4}$ of the unexpanded result through $\xi=100$.}
\label{fig:analytic}
\end{figure*}

Figure~\ref{fig:analytic} compares the exact angular dependence with the two expansions across the field range of interest. For $10^{-3}\le\xi\le100$ and $0\le\theta\le\pi/2$, the largest absolute residual of $\Aang_1$ is $3.75\times10^{-3}$, whereas $\Aang_2$ reduces it to $2.72\times10^{-4}$. Both extrema occur at $\xi=100$ near the magnetic-field direction. At that field the second-order residual is only about $0.5\%$ of the angular correction itself. The main value of $\Aang_1$ and $\Aang_2$ is therefore interpretive: they expose the sign, scaling, and angular morphology of the non-separable response and provide compact benchmarks for independent calculations. When $\kp$ and $\ks$ are already available numerically, the exact rationalized expression in Eq.~(\ref{eq:rational}) is equally inexpensive and should be used directly.

\begin{figure}[!t]
\centering
\includegraphics[width=\columnwidth]{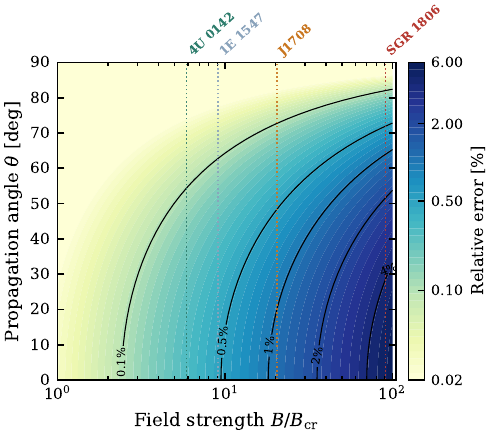}
\caption{Exact-normalized factorization error $100[\Aang^{-1}(\xi,\theta)-1]$ over the field--angle plane. The error vanishes at perpendicular propagation and approaches its maximum toward the field direction. Dotted vertical lines show the polar fields $B_p=2B_d$ adopted in Table~\ref{tab:sources}.}
\label{fig:map}
\end{figure}

\section{Domain of validity}\label{sec:accuracy}
The angular correction derived above comes from retaining the nonlinear algebraic form of the local constitutive relations instead of expanding the refractive indices strictly to first order in the response. This distinction separates a structural result from its one-loop numerical realization. Within the same local effective-Lagrangian framework, higher-loop terms modify the field derivatives that enter $\kp$ and $\ks$, while the common angular form of the two local indices in Eq.~(\ref{eq:common}) is retained. The positive-series argument and its monotonicity conclusion therefore continue to apply whenever the corrected coefficients remain on the same positive-birefringence branch.

The numerical size quoted in this paper is nevertheless only one-loop accurate. The leading birefringence is $\mathcal O(\alpha)$, whereas the absolute difference between the unexpanded and factorized forms begins formally at $\mathcal O(\alpha^2)$. Genuine two-loop contributions therefore enter at the same formal order as some products retained in the unexpanded one-loop expression. The two-loop Heisenberg--Euler action contains both the usual one-particle-irreducible contribution \cite{Ritus1975} and a finite one-particle-reducible term even for constant backgrounds \cite{GiesKarbstein2017}; the latter can become important in strong-field higher-loop expansions \cite{Karbstein2019,Karbstein2019Erratum}. We therefore interpret the theorem as a robust statement about the local angular structure, but the numerical error values as predictions of the unexpanded one-loop constitutive model rather than as a complete $\mathcal O(\alpha^2)$ QED result.

A second limitation is photon momentum. The local Heisenberg--Euler description applies to soft photons in a background varying slowly on the electron Compton scale. In the IXPE band, $\omega/m_e\lesssim1.6\times10^{-2}$, but the strong-field polarization tensor depends separately on the momentum components parallel and perpendicular to the external field \cite{Karbstein2013,HattoriItakuraI}. The combination $\chi_\gamma=(\omega/m_e)\xi\sin\theta$ is therefore useful as a kinematic indicator of transverse momentum, but it is not by itself a quantitative error estimate for $\dn$. The near-axis region is kinematically favourable for the local approximation because the transverse momentum vanishes with $\sin\theta$; a full momentum-dependent calculation would nevertheless be required for a systematic dispersive correction.

The zero-temperature approximation is also appropriate at the accuracy targeted here. Persistent magnetar soft X-ray spectra typically contain thermal components with $kT$ of a few tenths of a keV \cite{OlausenKaspi2014}, so that $kT/(m_ec^2)\sim10^{-3}\ll1$. In this low-temperature regime the leading thermal correction to the Heisenberg--Euler Lagrangian first appears at two loops. Recent work has shown that it can be generated from derivatives of the zero-temperature one-loop action and dressed by one-particle-reducible tadpole structures \cite{Karbstein2026}. Such terms belong to a higher-order precision budget; they do not alter the interpretation of the present calculation as a zero-temperature one-loop angular correction.

\section{Numerical validation}\label{sec:numerics}
We evaluate the one-loop response over $10^{-3}\le\xi\le100$ using arbitrary precision, extending the field range of Paper~I \cite{PaperI} so that the polar field of the strongest illustrative magnetar is covered directly. The implementation reproduces the weak-field limits, the strong-field asymptotes, the positive-birefringence branch, and the values reported in Paper~I. As an independent check, we also evaluate $\gs$, $q=\gs\kp$, and $m_v=-\gs\ks$ by Schwinger proper-time quadrature, switching to a small-proper-time Taylor expansion where direct evaluation would suffer from cancellation. At the eight benchmark fields $\xi=10^{-3},10^{-2},0.1,1,3,10,45.3,100$, the two implementations agree to much better than $10^{-12}$ in relative precision. This agreement provides an independent check of the one-loop coefficients themselves; it does not test higher-loop, dispersive, or thermal physics that lies outside the model.

For the angular splitting we use the rationalized form in Eq.~(\ref{eq:rational}) rather than subtracting nearly equal refractive indices near the field axis. We also checked the stability of the dipole integrations by increasing both the radial and surface quadrature resolutions; the source values quoted below are stable at better than $2\times10^{-4}$ in relative terms. With converged quadrature, increasing the radial upper limit from $60\RNS$ to $300\RNS$ changes the phase-error ratio by less than $6\times10^{-5}$ relative for all four illustrative source fields.

\begin{figure*}[!t]
\centering
\includegraphics[width=0.94\textwidth]{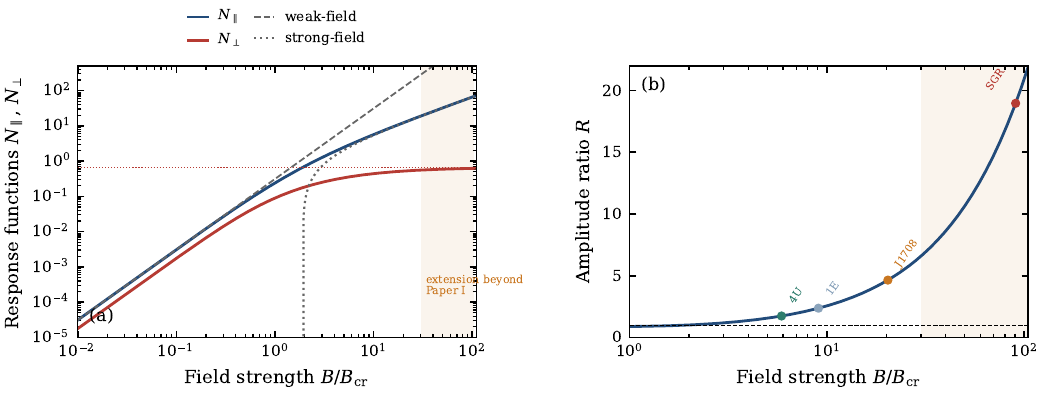}
\caption{Finite-field coefficient validation. \textbf{(a)} $\Npar$ and $\Nperp$ with weak- and strong-field limits; the shaded region is the extension beyond the $\xi\le30$ range used in Paper~I. \textbf{(b)} Ratio $R=\dn_{\rm wf}/\dn_{\rm exact}$ at perpendicular propagation. Source markers use the polar fields $B_p=2B_d$, for which SGR~1806$-$20 reaches $\xi_p\simeq90.6$.}
\label{fig:coeff}
\end{figure*}

\begin{table*}[!t]
\centering\small
\caption{Finite-field one-loop coefficients. $\dn$ and $R=\dn_{\rm wf}/\dn_{\rm exact}$ are evaluated at $\theta=\pi/2$. The ratio $R$ measures the finite-field change in the perpendicular-propagation amplitude relative to the weak-field result, whereas $\dmax$ is the maximum factorization error normalized to the exact splitting, as defined in Eq.~(\ref{eq:dmax}).}
\label{tab:coeff}
\setlength{\tabcolsep}{8pt}
\begin{tabular}{rrrrrrr}
\toprule
$\xi$ & $\Npar$ & $\Nperp$ & $\dn$ & $R$ & $\dmax$ [\%] & $\gs-1$\\
\midrule
0.1  & 0.003095 & 0.001748 & $7.82\!\times\!10^{-7}$ & 0.990 & 0.00042 & $-1.02\!\times\!10^{-6}$\\
1    & 0.23964  & 0.08945  & $8.72\!\times\!10^{-5}$ & 0.888 & 0.0287 & $-6.90\!\times\!10^{-5}$\\
3    & 1.2670   & 0.25071  & $5.90\!\times\!10^{-4}$ & 1.181 & 0.132 & $-2.79\!\times\!10^{-4}$\\
10   & 5.6625   & 0.44127  & $3.03\!\times\!10^{-3}$ & 2.556 & 0.532 & $-7.68\!\times\!10^{-4}$\\
20   & 12.228   & 0.52167  & $6.78\!\times\!10^{-3}$ & 4.567 & 1.110 & $-1.16\!\times\!10^{-3}$\\
30   & 18.852   & 0.55702  & $1.06\!\times\!10^{-2}$ & 6.587 & 1.688 & $-1.41\!\times\!10^{-3}$\\
45.3 & 29.019   & 0.58520  & $1.64\!\times\!10^{-2}$ & 9.688 & 2.570 & $-1.69\!\times\!10^{-3}$\\
60   & 38.802   & 0.60056  & $2.20\!\times\!10^{-2}$ & 12.683 & 3.416 & $-1.88\!\times\!10^{-3}$\\
80   & 52.120   & 0.61355  & $2.95\!\times\!10^{-2}$ & 16.783 & 4.561 & $-2.08\!\times\!10^{-3}$\\
100  & 65.444   & 0.62196  & $3.70\!\times\!10^{-2}$ & 20.910 & 5.701 & $-2.24\!\times\!10^{-3}$\\
\bottomrule
\end{tabular}
\end{table*}

\section{Magnetar dipole applications}\label{sec:dipole}
\subsection{Dipole geometry}
To translate the local QED correction into an astrophysical scale, the dipole-field convention must be fixed explicitly. We take the standard characteristic spin-down estimate
\begin{equation}
B_d=3.2\times10^{19}(P\dot P)^{1/2}\ \mathrm{G}
\label{eq:Bd}
\end{equation}
as the equatorial normalization of an ideal centred dipole. The associated polar field is therefore $B_p=2B_d$. With this convention, the surface-field magnitude and the angle between a radial ray and the local magnetic field are
\begin{align}
B(\theta_m)&=\frac{B_p}{2}\sqrt{1+3\cos^2\theta_m}\nonumber\\
&=B_d\sqrt{1+3\cos^2\theta_m},\\
\tan\theta_B&=\frac12\tan\theta_m.
\label{eq:dipole}
\end{align}
This distinction matters directly for the finite-field correction. For SGR~1806$-$20, $B_d\simeq2\times10^{15}$~G corresponds to $B_p\simeq4\times10^{15}$~G and hence $\xi_p=90.6$, not $45.3$. Extending the finite-field coefficients to $\xi=100$ therefore covers the full polar field in this illustrative centred-dipole model.

\subsection{Surface-averaged correction}
\begin{figure*}[!t]
\centering
\includegraphics[width=0.94\textwidth]{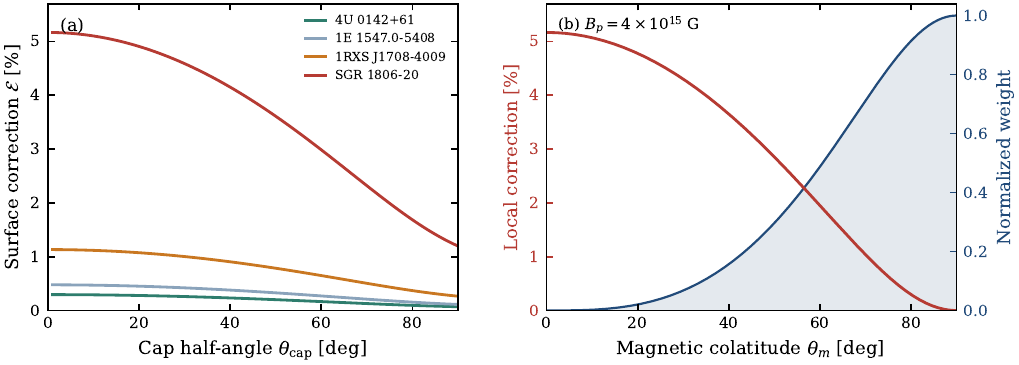}
\caption{Surface integration for a centred dipole. \textbf{(a)} Cap-integrated factorization error for the four dipole normalizations in Table~\ref{tab:sources}, assuming uniform surface emission. \textbf{(b)} For the SGR-class polar field, the local fractional error rises toward the pole while the birefringent weight $\dn\sin\theta_m$ vanishes there, suppressing the contribution of the formal near-axis maximum to the surface integral.}
\label{fig:cap}
\end{figure*}

For an emitting cap of half-angle $\theta_{\rm cap}$ we define
\begin{equation}
\mathcal E(\theta_{\rm cap})=
\frac{\int_0^{\theta_{\rm cap}}|\dn_{\rm exact}-\dn_{\rm fact}|\,w(\theta_m)\,d\theta_m}
{\int_0^{\theta_{\rm cap}}\dn_{\rm exact}\,w(\theta_m)\,d\theta_m},
\label{eq:cap}
\end{equation}
where $\dn_{\rm fact}=\dn(\xi,\pi/2)\sin^2\theta_B$. Figure~\ref{fig:cap} shows the result for the four field normalizations listed in Table~\ref{tab:sources}. The local fractional correction grows toward the magnetic pole, but the absolute birefringence and the surface-area weight both vanish there. Surface averaging therefore shifts the physical weight away from the formal near-axis maximum and samples a broader range of colatitudes.

For a $20^\circ$ cap with uniform surface weighting, $\mathcal E$ is $0.279\%$, $0.452\%$, $1.074\%$, and $4.907\%$ for 4U~0142$+$61, 1E~1547.0$-$5408, 1RXS~J1708$-$4009, and SGR~1806$-$20, respectively. The small-cap result is insensitive to simple changes in the weighting: replacing $w=\sin\theta_m$ by $w\propto\sin\theta_m\cos\theta_m$ or $w\propto\sin\theta_m B^2$ changes $\mathcal E(20^\circ)$ by at most $0.002$ percentage points, corresponding to less than $0.05\%$ of its value. This robustness is local to compact emitting regions. For a hemisphere with $B_p=4\times10^{15}$~G, the same three prescriptions give $1.20\%$, $2.01\%$, and $1.65\%$, respectively.

\subsection{Phase accumulation}
Along a radial ray, $\theta_B$ is constant for a centred dipole while $B\propto r^{-3}$. The accumulated vacuum phase is
\begin{equation}
\Phi=\omega\int_{\RNS}^{r_{\max}}\dn[\xi(r),\theta_B]dr,
\end{equation}
and we define
\begin{equation}
\delta_\Phi\equiv\frac{\Phi_{\rm fact}-\Phi_{\rm exact}}{\Phi_{\rm exact}}.
\label{eq:dphi}
\end{equation}
Within the nondispersive model the common factor $\omega$ cancels from this ratio. We integrate to $60\RNS$; the convergence test in Sect.~\ref{sec:numerics} shows that extending the upper limit has a negligible effect on the quoted phase-error ratios.

For the near-axis reference ray $\theta_m=5^\circ$ ($\theta_B=2.51^\circ$), the four phase errors are $0.154\%$, $0.245\%$, $0.554\%$, and $2.301\%$. They are smaller than the corresponding surface corrections because a radial ray rapidly leaves the supercritical-field region as $B\propto r^{-3}$; most of the path is therefore accumulated where the angular correction is already weak. These values refer only to the accumulated vacuum phase. Predicting an observed polarization angle or degree requires the coupled plasma--vacuum eigenmodes, possible mode conversion, ray geometry, and the emission pattern \cite{LaiHo2002,LaiHo2003,vAL2006,DinhThi2026}.

\begin{figure*}[!t]
\centering
\includegraphics[width=0.94\textwidth]{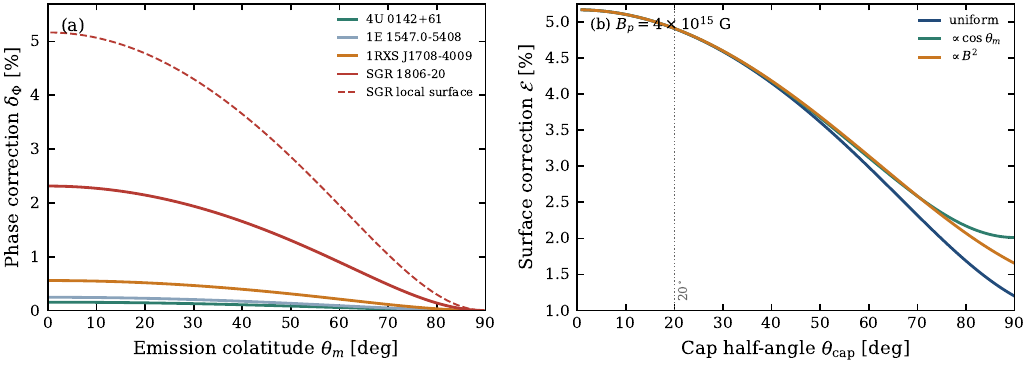}
\caption{Integrated diagnostics. \textbf{(a)} Radial accumulated-phase error versus emission colatitude for the four source dipole normalizations. The dashed curve is the local surface error for the SGR field and illustrates the geometric dilution along the ray. \textbf{(b)} Dependence of the SGR-class cap-integrated error on the emitting-cap size for three representative surface weightings. The curves are essentially indistinguishable for small caps, including $20^\circ$, but separate for broad emitting regions; at a hemisphere they give $1.20\%$, $2.01\%$, and $1.65\%$, respectively.}
\label{fig:phase}
\end{figure*}

\begin{table*}[!t]
\centering\small
\caption{Illustrative magnetar values using standard spin-down dipole strengths $B_d$ \cite{OlausenKaspi2014}. For the centred-dipole surface calculation we use $B_p=2B_d$. $\dmax$ is the exact-normalized near-axis error at the pole, and $\ell_B$ is evaluated there at perpendicular propagation; $\mathcal E$ is for a $20^\circ$ cap with uniform weighting; $\delta_\Phi$ is for the radial ray from $\theta_m=5^\circ$; $\ell_B=(\omega\dn)^{-1}$ is quoted at 4~keV and $\theta=\pi/2$.}
\label{tab:sources}
\setlength{\tabcolsep}{7pt}
\begin{tabular}{lrrrrrrr}
\toprule
Source & $B_d$ [$10^{14}$G] & $B_p$ [$10^{14}$G] & $\xi_p$ & $\dmax$ [\%] & $\mathcal E$ [\%] & $\delta_\Phi$ [\%] & $\ell_B$ [cm]\\
\midrule
4U 0142$+$61      & 1.3  & 2.6  & 5.89  & 0.295 & 0.279 & 0.154 & $3.19\!\times\!10^{-6}$\\
1E 1547.0$-$5408  & 2.0  & 4.0  & 9.06  & 0.478 & 0.452 & 0.245 & $1.84\!\times\!10^{-6}$\\
1RXS J1708$-$4009 & 4.5  & 9.0  & 20.39 & 1.132 & 1.074 & 0.554 & $7.12\!\times\!10^{-7}$\\
SGR 1806$-$20     & 20.0 & 40.0 & 90.62 & 5.167 & 4.907 & 2.301 & $1.47\!\times\!10^{-7}$\\
\bottomrule
\end{tabular}
\end{table*}

\subsection{Physical implications}
The hierarchy between local, surface-averaged, and path-integrated corrections is physically more informative than the near-axis maximum by itself. Close to the magnetic axis the relative factorization error is largest, but $\dn\rightarrow0$ and the two polarization eigenmodes become degenerate. The region that contributes most strongly to an integrated signal is therefore displaced from the axis. With the common exact normalization adopted here, the SGR~1806$-$20 near-axis error is $5.17\%$, while the $20^\circ$ cap average is $4.91\%$ and the representative radial phase error is $2.30\%$. Surface averaging therefore produces a modest but genuine suppression, and radial propagation reduces the correction further because the dipole field rapidly falls below $\Bcr$.

The observable polarization is set on a still larger scale. In the standard adiabatic picture, the photon polarization follows the local vacuum eigenmodes until the polarization-limiting radius, where mode tracking ceases and the polarization direction freezes \cite{HeylShaviv2003}. Paper~I found that finite-field corrections to the birefringence amplitude leave this radius unchanged to better than $10^{-12}$ in the same centred-dipole model, because decoupling occurs at $r_{\rm pl}\sim10^2\RNS$, where $B\ll\Bcr$ \cite{PaperI}. The angular correction studied here has the same weak-field limit and is therefore likewise concentrated well inside the freeze-out region. The birefringence lengths in Table~\ref{tab:sources}, $\ell_B\sim10^{-7}$--$10^{-6}$~cm at 4~keV near the surface, are many orders of magnitude shorter than the magnetic-field variation scale. The modes are thus strongly adiabatic where the local angular correction is largest. In a smooth centred dipole, the several-percent near-surface correction should not be interpreted as a comparable shift in the final observed polarization angle or degree; its clearest role is in the local eigenvalue splitting and the phase accumulated through the supercritical region.

The case of 1E~1547.0$-$5408 is useful in this respect because recent phase- and energy-resolved X-ray polarimetry provides a direct observational context for magnetar polarization transport \cite{StewartNature2026}. In the simplified centred-dipole calculation used here, its angular-factorization correction remains sub-percent: $\dmax=0.478\%$, $\mathcal E(20^\circ)=0.452\%$, and $\delta_\Phi=0.245\%$. These values quantify one vacuum-sector approximation entering a more complete radiative-transfer and mode-evolution problem rather than a predicted fractional change in the measured polarization.

The distinction becomes especially important near a plasma--vacuum resonance, where the vacuum and plasma terms are comparable. A small change in the vacuum response can then alter the combined eigenvectors and resonance structure in a way that cannot be represented by multiplying a vacuum-only conversion probability by the same fractional correction. Applications to mode conversion must therefore be formulated at the level of the full dielectric response. The present calculation is intended instead as a controlled correction to the vacuum eigenvalue splitting and to phase accumulation.

\section{Conclusions}\label{sec:conclusion}
This work extends our previous finite-field study of QED vacuum birefringence \cite{PaperI} from the field-strength dependence of the splitting to its angular dependence. The two effects are distinct. Retaining the correct finite-field amplitude does not, by itself, make the familiar $\sin^2\theta$ angular factor exact.

Within the local one-loop Heisenberg--Euler model, the arbitrary-angle refractive indices can be written in a common form that exposes the residual angular dependence analytically. On the positive-birefringence branch considered here, the usual factorized prescription overestimates the exact finite-field splitting at every non-perpendicular angle. The relative correction is largest for propagation close to the magnetic-field direction, but the absolute birefringence vanishes there. The first- and second-order expansions make the sign, scaling, and angular morphology explicit; the latter follows the full angular profile to better than $2.72\times10^{-4}$ in the normalized splitting over $B/B_{\rm cr}\le100$. These expansions are best viewed as analytic diagnostics and benchmarks. When the finite-field coefficients are already available, the exact rationalized expression is just as inexpensive and should be used directly.

The magnetar calculations show where this local correction is physically relevant. For moderate magnetar fields it is sub-percent, while for the strongest example considered here the exact-normalized near-axis error reaches $5.17\%$. Surface averaging reduces this to $4.91\%$ for a $20^\circ$ cap, and radial phase accumulation reduces it further to $2.30\%$ along the representative near-axis ray. In the same centred-dipole setting, however, the polarization freezes much farther out, at radii of order $10^2\RNS$ where $B\ll\Bcr$ \cite{PaperI,HeylShaviv2003}. The correction identified here is therefore primarily a near-surface strong-field effect, not a several-percent prediction for the final observed polarization.

The scope of the result should remain explicit. The analytic sign and monotonicity follow from the local constitutive angular structure on the positive-birefringence branch, whereas the numerical values in this paper use one-loop QED coefficients. Genuine two-loop terms enter at the same formal perturbative order as part of the retained algebraic nonlinearity, and momentum-dependent and thermal effects are omitted. Observable polarization additionally requires plasma physics, emission geometry, magnetic topology, and mode evolution. Within these limits, the paper provides a controlled characterization of the angular error that remains after the finite-field amplitude correction of Paper~I has already been included.

\section*{Acknowledgements}

\section*{Data and code availability}
The complete manuscript source and reproducibility notebook are available with the submission. The notebook includes all calculations, validation tests, and executable cells for each figure.

\end{document}